\documentclass[
    ,final            
  ]
  {aipproc}

\layoutstyle{8x11single}

\usepackage{graphicx}

\newcommand*{\bC}{\ensuremath{\mathbf{C}}}

\newcommand{\be}{\begin{equation}}
\newcommand{\ee}{\end{equation}}
\newcommand{\bes}{\begin{equation*}}
\newcommand{\ees}{\end{equation*}}
\newcommand{\bea}{\begin{eqnarray}}
\newcommand{\eea}{\end{eqnarray}}
\newcommand{\beas}{\begin{eqnarray*}}
\newcommand{\eeas}{\end{eqnarray*}}

\begin{document}

\title{A New Approach for Delta Form Factors}


\author{C.\ Aubin}{
  address={Dept.\ of Physics, Fordham University, Bronx NY},
  altaddress={Dept.\ of Physics, College of William and Mary,
  Williamsburg, VA}
}

\author{K.\ Orginos}{
  address={Dept.\ of Physics, College of William and Mary,
  Williamsburg, VA},
  altaddress={Thomas Jefferson National Accelerator Facility, Newport News, VA }
}

\begin{abstract}

We discuss a new approach to reducing excited state contributions from two- and three-point correlation functions in lattice simulations. For the purposes of this talk, we focus on the $\Delta(1232)$ resonance and discuss how this new method reduces excited state contamination from two-point functions and mention how this will be applied to three-point functions to extract hadronic form factors.

\end{abstract}

\maketitle


The calculation of the electromagnetic form factors for mesons and baryons is crucial to understanding the structure of hadronic states in QCD. However, they are notoriously difficult to both measure experimentally as well as calculate theoretically due to the complications that arise from the strong interactions. In the case of the $\Delta(1232)$ resonance, the form factors themselves are not currently experimentally accessible, although two of the form factors in the static limit are known (the charge) or measured to some degree (the magnetic dipole moment). For the nucleon, experimental results do exist for the form factors as a function of the momentum transfer. This makes a calculation of the nucleon form factors both a check of methodology as well as of QCD, and allows us to be confident that lattice results for the $\Delta$ form factors are reasonable.

The electromagnetic form factors of the $\Delta$ are encoded in the matrix element
\be
	\left\langle\Delta(p')| J_\mu | \Delta(p)\right\rangle
	=
	\bar{u}_\alpha(p') \Gamma_{\alpha\beta\mu} u_\beta(p)
\ee
where $u_\alpha$ is a Rarita-Schwinger vector-spinor describing the external $\Delta$, and $J_\mu = \sum_q \bar{q}\gamma_\mu q$ is the vector current. The Lorentz structure of $\Gamma$ is given, for example, in \cite{Pascalutsa:2006up}, and has four form factors $F_{1,2,3,4}^*(Q^2)$ that are functions of $Q^2 = -(p'-p)^2$ alone. These form factors give rise, in the limit $Q^2\to0$, to the electric charge, magnetic dipole moment, electric quadrupole moment, and magnetic octupole moment of the $\Delta$. 

Of these moments, the charge is of course known, and from the PDG \cite{PDG}, we have
\bea
	\mu_{\Delta^{++}} & = & (5.6\pm 1.9)\mu_N\ ,\\
	\mu_{\Delta^+} & = & (2.7\pm 3.5)\mu_N \ ,
\eea
where we have added all of the errors (including theoretical) in quadrature just to get an idea for how well these are determined experimentally. Thus, it is essential even for this simple quantity to have a well-determined lattice result. Some unquenched results were obtained using the form factor approach in \cite{Alexandrou:2008bn} and using a background field technique \cite{Aubin:2008qp}, but there are several difficulties that arise in these different methods. For this work, we will focus on the difficulties with the form factor approach. 

These are determined by calculating the 3-point correlator:
\be
	C^{3\rm pt}(t_i,t,t_f,\mathbf{p}_i,\mathbf{p}_f)
	=
	FT\left[\left\langle 0|\chi(t_f,\mathbf{x}_f) J_\mu(t,\mathbf{0})
	\overline{\chi}(t_i,\mathbf{x}_i)|0\right\rangle\right]\ ,
\ee
where $FT$ is the Fourier Transform of the correlator, and $\chi$ is some appropriate interpolating operator for the $\Delta$. In the large time limit $t_f\gg t\gg t_i$:
\be
	C^{3\rm pt}(t_i,t,t_f,\mathbf{p}_i,\mathbf{p}_f)
	\to
	Z(\mathbf{p}_i,\mathbf{p}_f) e^{-E_f(t_f-t)}e^{-E_i(t-t_i)}
	\left\langle \Delta(p_f)|J_\mu(0)| \Delta(p_i)\right\rangle
	+\cdots\ .
\ee
Here we have schematically written this so that $Z$ contains various overlap factors of the form $\langle 0|\chi|\Delta\rangle$ as well as other kinematic factors that are known. The dots denote contributions from excited states that are generally ignored. The standard approach then is to note that the overlap factors and kinematics can be canceled by an appropriate ratio with 2-point correlators, and from these ratios we can extract the matrix element of interest. Note that additional complications arise in separating the $\Delta$ state and the $N-\pi$ state when the pion mass is below 300 MeV. Currently we are not near this regime, so it is not an issue for the current analysis.

\begin{figure}[tbp]
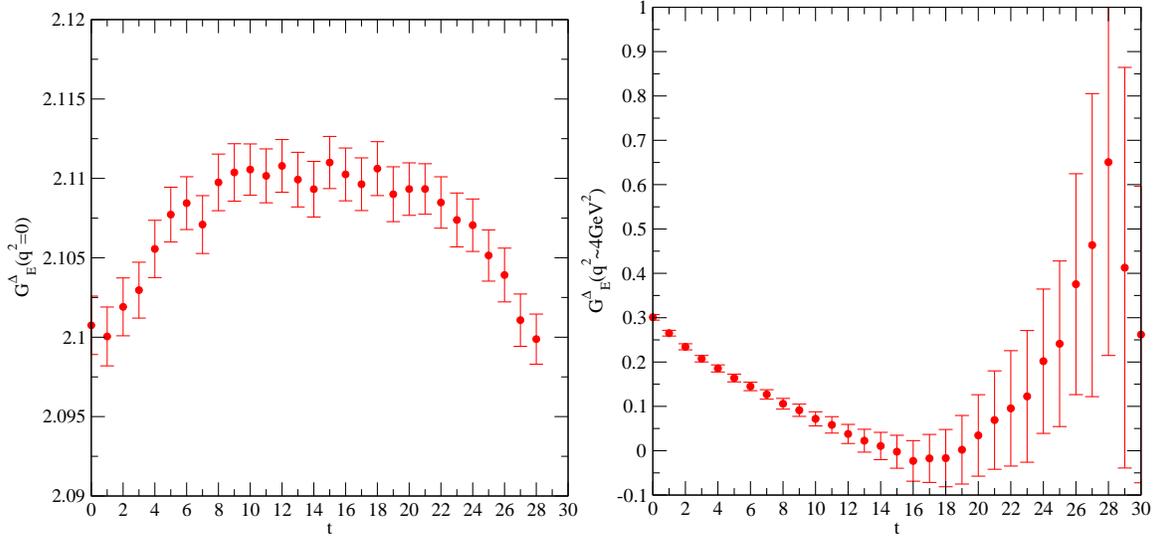

\includegraphics[width=3in]{GE_Delta_p_0.eps}\hfill
\includegraphics[width=3in]{GE_Delta_p_11.eps}
\caption{On the left is the matrix element $G_E$ for the $\Delta^{++}$ evaluated at $q^2=0$. While there is significant excited state contamination, there is a noticeable plateau that gives (up to renormalization) the charge of the $\Delta$. On the right is the same matrix element for $q^2 = 11(2\pi/a_s L)^2\approx 4$ GeV${}^2$, where there is no trustworthy signal.}
\label{fig:plateau}
\end{figure}

The problem is that the contamination from excited states can be seen to be large, even for the simplest cases. Take here the $\Delta^{++}$ $E0$ form factor, $G_{E0}(q^2)$, which is a linear combination of the $F^*_i$ form factors. In the limit $q^2\to0$, $G_E(0)= +2$, the electric charge of the $\Delta^{++}$ in units of $|e|$. We show on the left of Fig.~\ref{fig:plateau} the appropriate ratio of 3-to-2-point correlators to get this quantity using the Hadron Spectrum collaboration lattices of 2+1-flavor anisotropic Clover lattices \cite{Lin:2008pr} (here with a volume of $16^3\times128$ and a pion mass of roughly 390 MeV, and $a_t^{-1}\approx 5.5\,{\rm GeV},a_s/a_t\approx 3.5$). The source and sink are located at $t=0,28$, and we can see a plateau in the center where we could reliably extract the form factor, but there is significant contamination from excited states, seen from the deviation from the plateau. 

This problem is more pronounced at higher momenta, and we show the same form factor at $q^2 = 11(2\pi/a_s L)^2\approx 4$ GeV${}^2$ on the right of Fig.~\ref{fig:plateau}. The location of a plateau is questionable at best for this plot.

We can extract the form factor using this standard approach and one gets reasonable results. For the $q^2=0$ point, we find the charge of the $\Delta^{++}$ is precisely twice that of the proton, and we can extract the renormalization factor for the vector current (note it is not conserved here because we are using a local current) from $G_{E0}(0)$, and we obtain
\[
	\frac{1}{Z_V} = 1.05(1)\ \mbox{(statistical errors only)}.
\]
While this determination is trustworthy, extracting the higher momenta form factors is dangerous, due to the excited state contamination. So we would like to examine a new approach, which makes use of the Variational method to better extract states that contribute to a correlator. For now we will discuss this in the context of 2-point correlators for simplicity. 

For a given state, there are many interpolating operators that could be used to calculate a two-point function, and one could form a matrix of correlators
\be
	C_{ij}(t) = \left\langle 0| O_i(t) \bar O_j(0)|0\right\rangle\ .
\ee
From this, by solving the generalized eigenvalue problem (GEVP)
\be
	C(t) \mathbf{x}
	=
	\lambda(t,t_0) C(t_0)\mathbf{x}\ ,
\ee
one can show (see \cite{Blossier:2009kd} and references therein) that the eigenvalues behave like
\be
	\lambda_i(t,t_0) \sim e^{-m_i(t-t_0)}+ \cdots\ .
\ee
The reference time $t_0$ is empirically chosen so that at that time, all $n$ states (for an $n\times n$ system) would contribute to the correlator; no more, no less. For our purposes, we find that the choice of $t_0$ has little effect on our results, primarily because we are using a very small basis of operators.

\begin{figure}[tbp]
\includegraphics[width=3in]{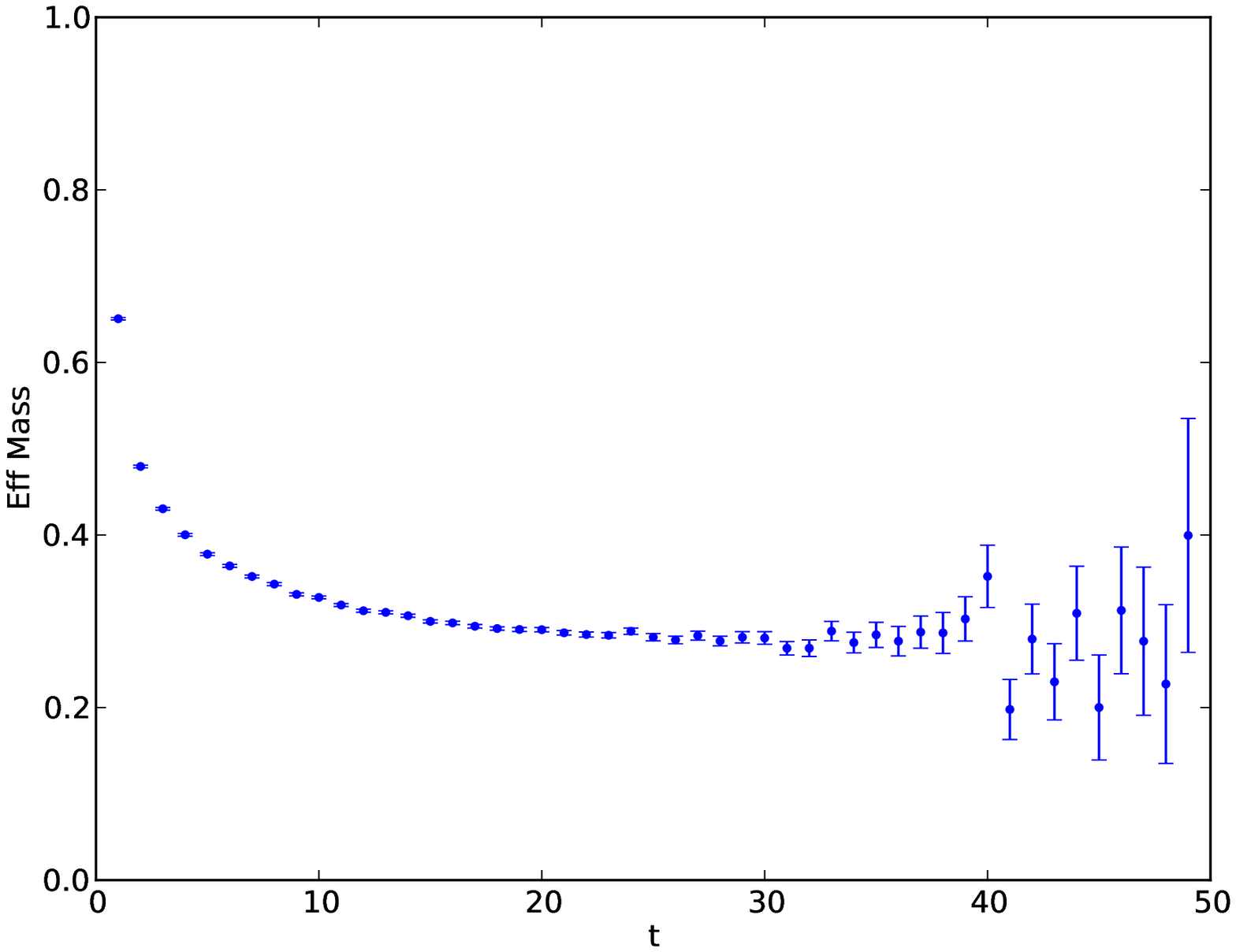}\hfill
\includegraphics[width=3in]{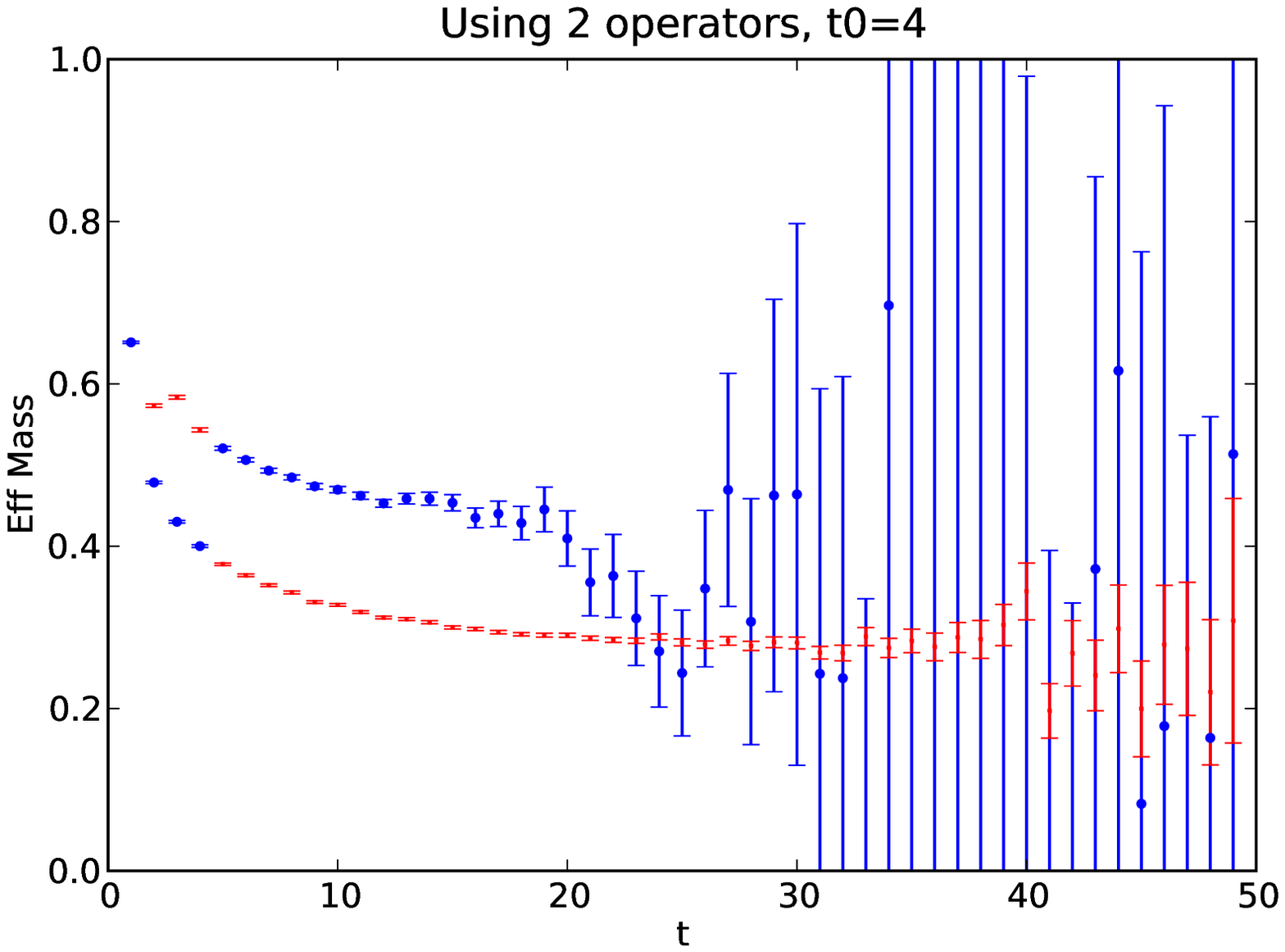}
\caption{On the left we show the effective mass for a single $\Delta$ operator with noticeable excited state contamination before reaching a plateau. On the right is the effective masses for the two states extracted using the GEVP and two different local $\Delta$ operators.}
\label{fig:noshift}
\end{figure}

If we restrict ourselves to local operators, there are two $\Delta$ interpolating fields we can use, and we can see that this has little effect to reduce contamination from excited states in the ground state, as shown by comparing the two plots in Fig.~\ref{fig:noshift}. On left is a single operator effective mass, and we can see the excited state contamination before the plateau. On the right is the case with two operators, and solving the GEVP. The ground state is unchanged, and although this allows us to perhaps extract an excited state, this is not what we are currently interested in.

So we would like to find a way to kill off the excited state contribution in the ground state. This is where the Generalized Pencil-of-Function (GPoF) method comes in \cite{Hua,Sarkar}. In a quantum mechanical system, the important point is that if $O_\Delta(t)$ is an interpolating operator for the $\Delta$, then so is
\be
	O^\tau_\Delta(t)
	\equiv
	e^{H\tau}
	O_\Delta(t)
	e^{-H\tau} = O_\Delta(t+\tau)
	\ ,
\ee
and this new operator is linearly independent from the original operator.

So if we use $O_\Delta(t),O^\tau_\Delta(t)$ as our two operators, we can construct a correlator matrix using only a single correlator. This matrix is
\bea
	\bC(t)
	 & = &
	\left(\begin{array}{cc}
	\langle O_\Delta(t) \bar{O}_\Delta(0)\rangle &
	\langle O^\tau_\Delta(t)\bar O_\Delta(0)\rangle \\
	\langle O_\Delta(t)\bar O^\tau_\Delta(0)\rangle &
	\langle O^\tau_\Delta(t)\bar O^\tau_\Delta(0)\rangle
	\end{array}\right)
	 =
	\left(\begin{array}{cc}
	\langle O_\Delta(t)\bar O_\Delta(0)\rangle &
	\langle O_\Delta(t+\tau)\bar O_\Delta(0)\rangle\\
	\langle O_\Delta(t)\bar O_\Delta(-\tau)\rangle &
	\langle O_\Delta(t+\tau)\bar O_\Delta(-\tau)\rangle
	\end{array}\right)\nonumber\\
	& = &
	\left(\begin{array}{cc}
	C(t) & C(t+\tau)\\
	C(t+\tau) & C(t+2\tau)
	\end{array}\right)\ .
\eea
We can replicate this and use a set of operators $O^{\tau,n}_\Delta(t) = O_\Delta(t+n\tau)$, and make this correlator as large as we wish. It turns out that using multiple shifts does not give us any additional information, so we will ultimately use just a single shift and a $\tau=4$. This would be something that is determined for each correlator one is interested in. Once this matrix is created then we follow through the same procedure as before with the GEVP. 

Note that when the correlator basis grows, there is more of a likelihood of zero singular values, so we perform an SVD cut on the correlator matrix to exclude states that have significantly smaller singular values than the largest by some cutoff, here chosen to be $\approx 10^{-3}$.

\begin{figure}[tbp]
\includegraphics[width=3in]{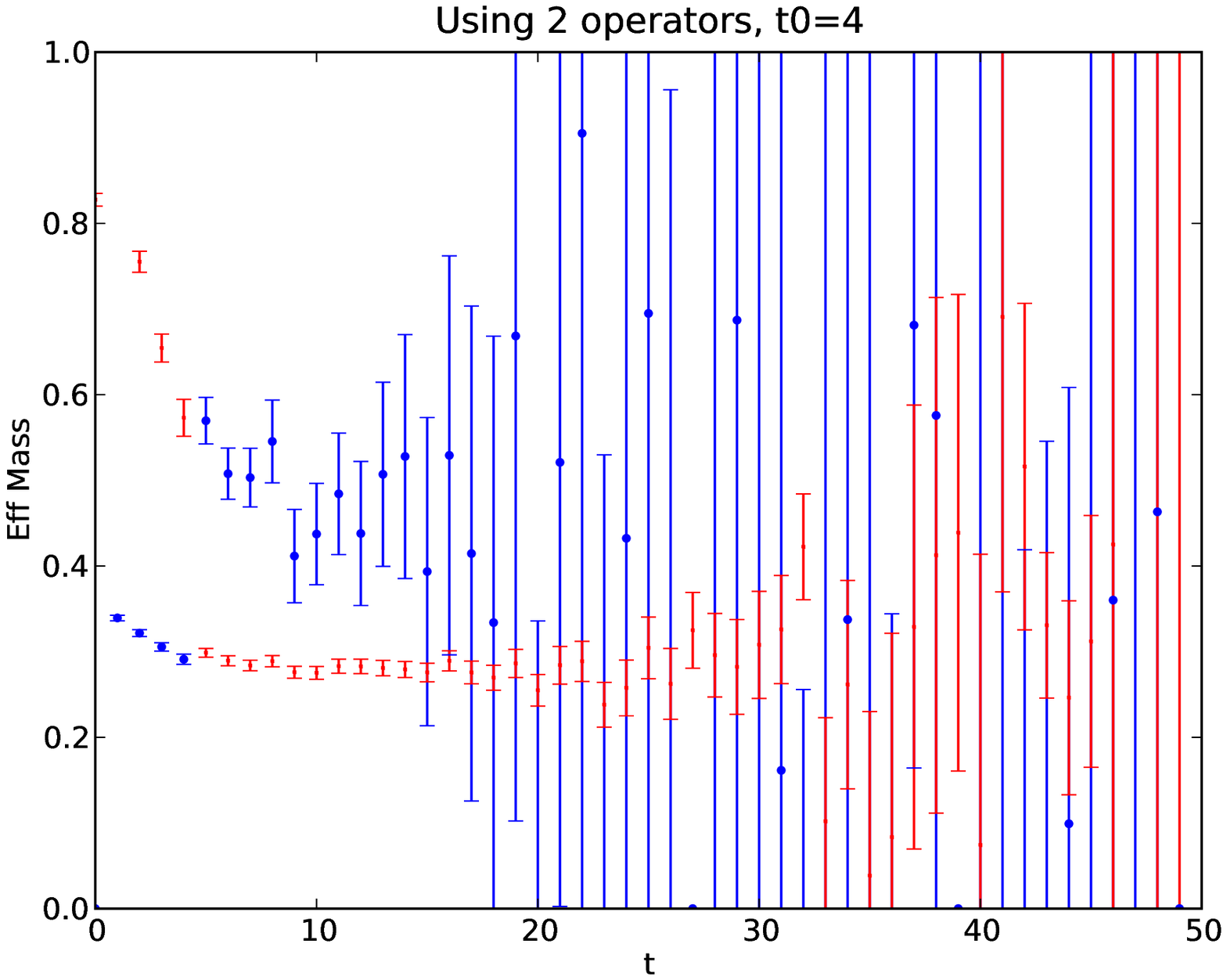}\hfill
\includegraphics[width=3in]{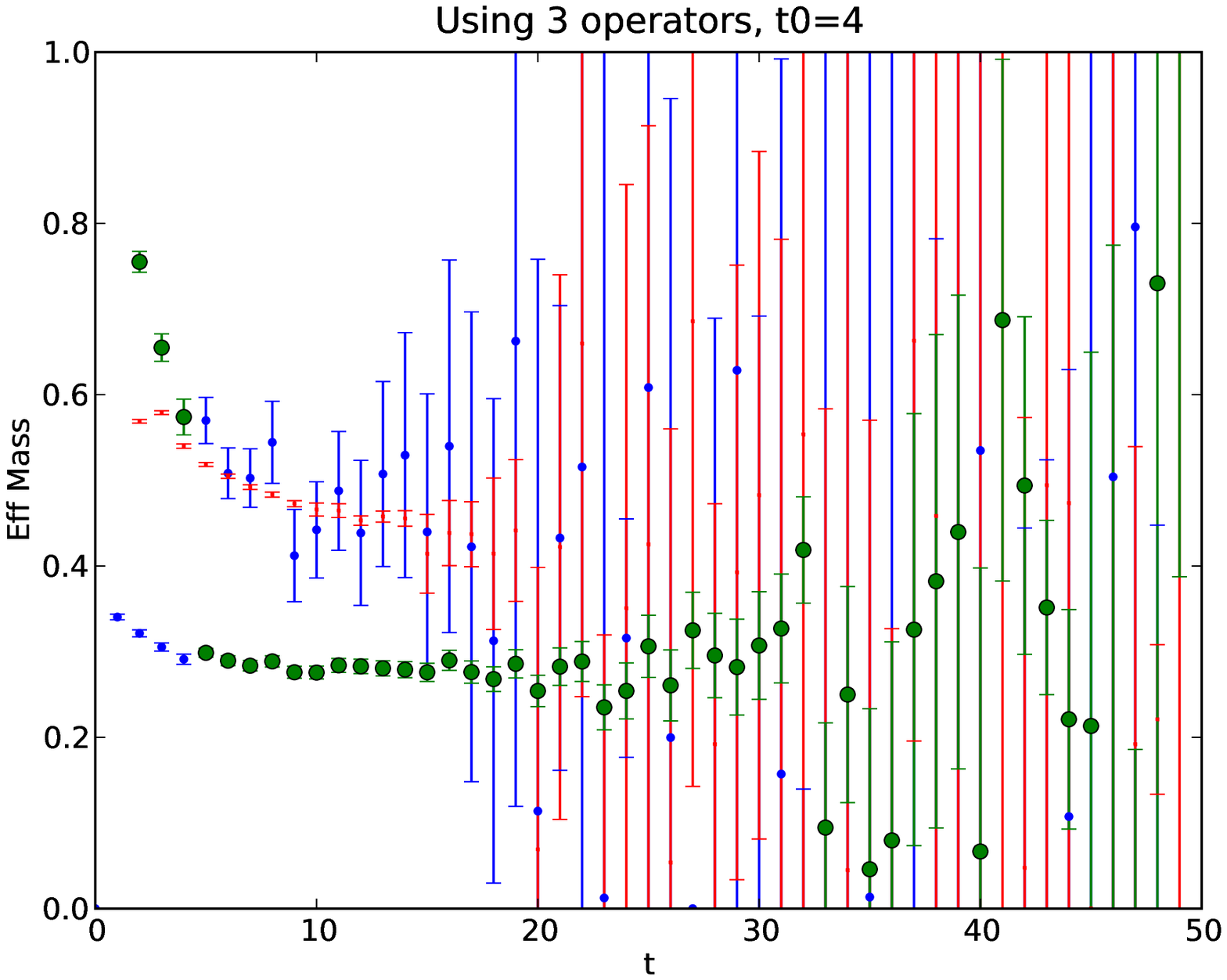}
\caption{This is the same as in Fig.~\ref{fig:noshift}, but with a single shift of $\tau=4$ for the operator(s) included.}
\label{fig:1shift}
\end{figure}

In Fig.~\ref{fig:1shift} we show the same operator(s) as before, but now using the GPoF method (again, $\tau=4$). On the left is one operator with one shift and the right uses both local operators and a single shift. We see the effective mass comes to a plateau much earlier than before, immediately after $t_0$ (which is also set to 4). Of course, the excited state(s) is(are) much noisier and we could not reliably extract information from that, but if one is only interested in the ground state, this is not an issue. However, one could apply the GPoF method to a large correlator basis to perhaps get cleaner signals for the lower lying states. On the right we perform the shift to the two operators, creating a $4\times4$ basis, but after the SVD cut, this is reduced to three operators. In either case, we see that the ground state is unchanged, and a plateau is reached far sooner than without the use of the GPoF method.

For the unshifted operator if we perform a correlated fit to the correlator using an exponential in the range $t\in\{29,40\}$, we get $a_tm_\Delta = 0.2770(50)$ with a confidence level of 78\%. For the single shift of $\tau=4$, we use the larger (and earlier) range $t\in\{5,28\}$, and get $a_tm_\Delta = 0.2825(17)$ with a confidence level of 84\%. Adding more shifts does not change this result, but it is amazing how well the signal improves with just this single shift.

It can be seen what is happening qualitatively, since the correlator $C(t+\tau)$ comes from the correlation with an operator that is $\tau$ time steps away from the other operator, and thus excited states are not going to contributed much to that, and even less to $C(t+2\tau)$. The ground state, however, is going to still contributed significantly to this correlator, so its signal, in a sense, is effectively enhanced.

In the three-point case, working through the correlator matrix, we would get
\bea
	\bC^{3-\rm pt}(t_i,t,t_f)
	& = &
	\left(\begin{array}{cc}
	C^{3-\rm pt}(t_i,t,t_f) & C^{3-\rm pt}(t_i,t,t_f+\tau)\\
	C^{3-\rm pt}(t_i,t+\tau,t_f+\tau) & C^{3-\rm pt}(t_i,t+\tau,t_f+2\tau)
	\end{array}\right)\ .
\eea
So unlike the two-point function, we have to actually use three different sink locations, and thus generate a factor of three in the computational cost. However, if this allows a better determination of our ground state signal, this should be worth the additional propagator generation.

Once the two-point correlator matrix is diagonalized with the vectors $\mathbf{x}$, we form a matrix $V_{ij} = (\mathbf{x}_j)_i$ and then use that to diagonalize the three-point correlator
\be
	\bC^{3-\rm pt}_{\rm diag}(t_i,t,t_f)
	=
	V^{-1}\bC^{3-\rm pt}(t_i,t,t_f)V\ .
\ee
Currently, the analysis using this approach is under way and we hope to see a noticeable improvement as we did in the two-point case.

This work was partially supported by the US Department of Energy, under contract nos. DE-AC05-06OR23177 (JSA), DE-FG02-07ER41527, and DE-FG02-04ER41302; and by the Jeffress Memorial Trust, grant J-813. 



\bibliographystyle{aipproc}   

\bibliography{writeup}

\begin{thebibliography}{8}
\expandafter\ifx\csname natexlab\endcsname\relax\def\natexlab#1{#1}\fi
\providecommand{\enquote}[1]{``#1''}
\expandafter\ifx\csname url\endcsname\relax
  \def\url#1{\texttt{#1}}\fi
\expandafter\ifx\csname urlprefix\endcsname\relax\def\urlprefix{URL }\fi
\providecommand{\eprint}[2][]{\url{#2}}

\bibitem[Pascalutsa et~al.(2007)]{Pascalutsa:2006up}
V.~Pascalutsa, M.~Vanderhaeghen, and S.~N. Yang, \emph{Phys. Rept.}
  \textbf{437}, 125--232 (2007).

\bibitem[Nakamura(2010)]{PDG}
K.~Nakamura, \emph{J. Phys.} \textbf{G37}, 075021 (2010).

\bibitem[Alexandrou et~al.(2009)]{Alexandrou:2008bn}
C.~Alexandrou, et~al., \emph{Phys. Rev.} \textbf{D79}, 014507 (2009),
  \eprint{0810.3976}.

\bibitem[Aubin et~al.(2009)]{Aubin:2008qp}
C.~Aubin, K.~Orginos, V.~Pascalutsa, and M.~Vanderhaeghen, \emph{Phys. Rev.}
  \textbf{D79}, 051502 (2009).

\bibitem[Lin et~al.(2009)]{Lin:2008pr}
H.-W. Lin, et~al., \emph{Phys. Rev.} \textbf{D79}, 034502 (2009).

\bibitem[Blossier et~al.(2009)]{Blossier:2009kd}
B.~Blossier, M.~Della~Morte, G.~von Hippel, T.~Mendes, and R.~Sommer,
  \emph{JHEP} \textbf{04}, 094 (2009).

\bibitem[Hua and Sarkar(1989)]{Hua}
Y.~Hua, and T.~Sarkar, \emph{IEEE transactions on antennas and propagation}
  \textbf{37}, 229--234 (1989).

\bibitem[Sarkar and Pereira(1995)]{Sarkar}
T.~Sarkar, and O.~Pereira, \emph{IEEE Antennas and Propagation Magazine}
  \textbf{37}, 48--55 (1995).

\end{thebibliography}

\end{document}